\documentclass{iaus}
\usepackage{graphicx}

\title[X-rays from Magnetic OB Stars] 
{X-rays from Magnetically Channeled Winds of OB Stars}

\author[David H. Cohen]   
{David H. Cohen$^1$}

\affiliation{$^1$Swarthmore College, Department of Physics and Astronomy, 500 College Ave., Swarthmore, Pennsylvania 19081}

\pubyear{2008}
\volume{250}  
\jname{Massive Stars as Cosmic Engines}
\editors{F. Bresolin, P.A. Crowther \& J. Puls, eds.}
\begin{document}

\maketitle

\begin{figure}[h]
\begin{center}
\vspace*{0.7in}
\hspace*{-0.25in}
 \includegraphics[scale=0.5,angle=0]{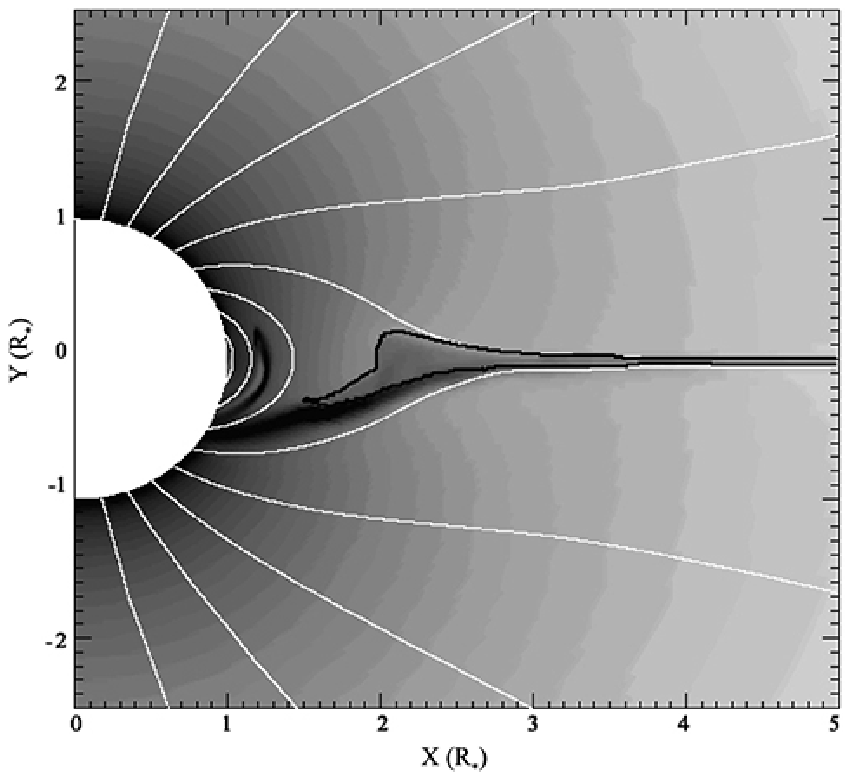} 
\vspace{2.in}
\hspace*{-1.8in}
 \includegraphics[scale=0.5,angle=0]{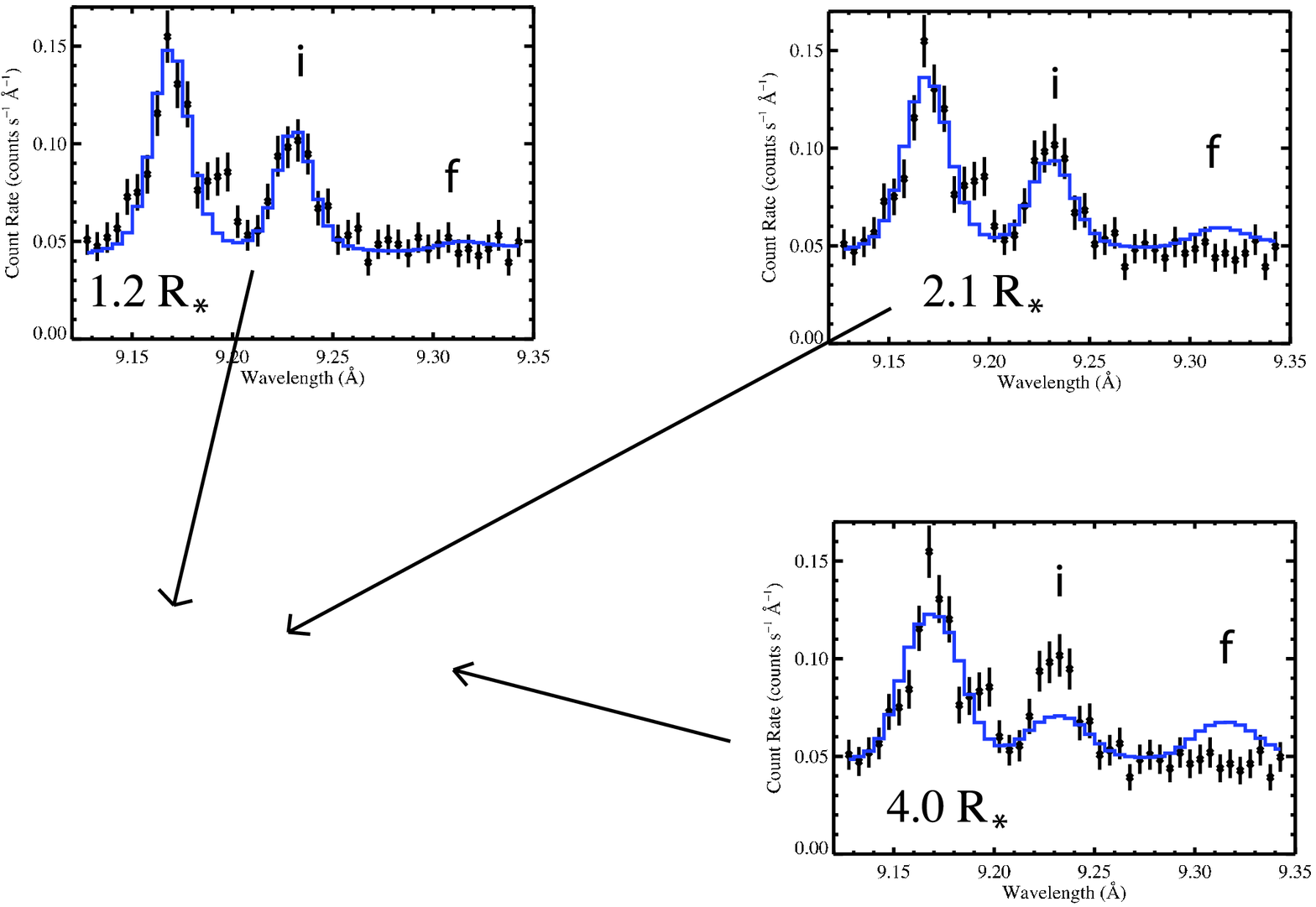} 
\vspace{-1.85in}
 \caption{A snapshot from a 2-D MHD simulation of $\theta^1~{\rm
     Ori~C}$, taken from \cite[Gagn\'{e} et al.\ (2005)]{Gagne2005},
   showing emission measure in grayscale, magnetic field lines as
   white contours, and with a thick, black contour enclosing plasma
   with temperature above $10^6$ K (lower left).  The three other
   panels show the {\it Chandra} spectrum in the vicinity of the
   helium-like Mg\, {\sc xi} complex, with a model of the resonance,
   intercombination ($i$), and forbidden ($f$) lines overplotted.  The
   relative strengths of the $f$ and $i$ lines are different in each
   of the three panels, as the three models were calculated assuming a
   source location of 1.2 ${\mathrm {R_{\ast}}}$, 2.1 ${\mathrm
     {R_{\ast}}}$, and 4.0 ${\mathrm {R_{\ast}}}$, respectively,
   starting at the top left and moving clockwise. The arrows indicate
   the approximate location in each case. The intermediate case -- $r
   = 2.1~{\mathrm {R_{\ast}}}$, which seems to agree with the MHD
   simulation -- is marginally consistent with the data (68\%
   confidence limit). }
   \label{fig:ftoi}
\end{center}
\end{figure}

OB stars with strong radiation-driven stellar winds and large-scale
magnetic fields generate strong and hard X-ray emission via the
Magnetically Channeled Wind Shock (MCWS) mechanism (\cite[Shore \&
Brown 1990; Babel \& Montmerle 1997; ud-Doula \& Owocki
2002]{sb1990,bm1997,uo2002}). There are four separate X-ray
diagnostics that confirm the MCWS scenario for the young, magnetized O
star that illuminates the Orion Nebula, $\theta^1~{\rm Ori~C}$, and
constrain the physical properties of its X-ray emitting magnetosphere:

\begin{description}

\item[1.~]{High X-ray temperatures, determined from thermal spectral
    model fitting.  The differential emission measure of
    $\theta^1~{\rm Ori~C}$ peaks at temperatures above 10 MK, which is
    in contrast to the few million K peak temperatures in mature,
    unmagnetized O stars (\cite[Wojdowski \& Schulz 2005]{ws2005}),
    and which is well reproduced by MHD simulations of the MCWS
    mechanism (\cite[Gagn\'{e} et al.\ 2005]{Gagne2005}).}

\item[2.~]{Relatively narrow X-ray emission lines.  The X-ray emitting
    plasma in the MCWS scenario is predominantly in the closed
    magnetic field regions and thus the plasma velocity is relatively
    low and the associated Doppler line broadening is modest.  This is
    seen in the MHD simulations and confirmed by the {\it Chandra}
    grating observations.}

\item[3.~]{The rotational modulation of the X-ray emission is
    consistent with part of the magnetosphere being eclipsed near
    phase 0.5, when the viewing orientation is magnetic equator-on.
    The depth of the eclipse provides information about the location
    of the X-ray emitting plasma (deeper eclipses imply more plasma
    close to the star).  The observed eclipse depth for $\theta^1~{\rm
      Ori~C}$ implies that the bulk of the plasma is within a stellar
    radius of the photosphere. This is a somewhat closer than the MHD
    simulations predict (e.g.\ \cite[Gagn\'{e} et al.\
    2005]{Gagne2005}).}

\item[4.~]{The ratio of the forbidden to intercombination line
    strengths in helium-like ions also puts a constraint on the
    location of the X-ray emitting plasma, via the sensitivity of
    these line ratios to the local UV mean intensity.  The closer the
    hot plasma is to the photosphere, the stronger the UV
    photoexcitation of electrons from the upper level of the forbidden
    line to the upper level of the intercombination line, and the
    smaller the $f/i$ ratio.  This is demonstrated in Fig.\
    \ref{fig:ftoi} for the Mg\, {\sc xi} complex in the co-added (over
    four observations) {\it Chandra} grating spectrum of
    $\theta^1~{\rm Ori~C}$, where we see that the very weak forbidden
    line requires a plasma location below $r \approx 2~{\mathrm
      R_{\ast}}$.}

\end{description}

\end{document}